# Detecting Change-Points in Time Series by Maximum Mean Discrepancy of Ordinal Pattern Distributions

**Mathieu Sinn**
IBM Research
Dublin, Ireland

**Ali Ghodsi**
Department of Statistics and
Actuarial Science, University of Waterloo
Waterloo, Ontario, Canada

**Karsten Keller**
Institute of Mathematics
University of Lübeck
Lübeck, Germany

## Abstract

As a new method for detecting change-points in high-resolution time series, we apply Maximum Mean Discrepancy to the distributions of ordinal patterns in different parts of a time series. The main advantage of this approach is its computational simplicity and robustness with respect to (non-linear) monotonic transformations, which makes it particularly well-suited for the analysis of long biophysical time series where the exact calibration of measurement devices is unknown or varies with time. We establish consistency of the method and evaluate its performance in simulation studies. Furthermore, we demonstrate the application to the analysis of electroencephalography (EEG) and electrocardiography (ECG) recordings.

## 1 INTRODUCTION

Detecting changes in the temporal evolution of a system (biological, physical, mechanical, etc.) is of major importance, e.g. in medical diagnostics, industrial quality control, or the analysis of financial markets. In statistics, the problem of inferring the time point of a change from a sequence of observations is known as change-point detection. There is a vast literature on parametric and non-parametric methods for testing the presence of change-points and for estimating their locations. For an overview, we refer to the monographs of Chen and Gupta (2000) and Brodsky and Darkhovsky (1993, 2000). Recently, Harchaoui et al. (2009) have introduced an estimator for change-points based on kernels.

In this paper, we propose to detect change-points by comparing the distributions of ordinal patterns in different parts of a time series. In order to capture differences among the distributions automatically, we apply the Maximum Mean Discrepancy (MMD) criterion recently introduced by Gretton et al. (2007a, 2007b). The proposed method is computationally fast and robust with respect to (non-linear) monotonic transformations of the observations, which makes it particularly attractive for the exploration of high-resolution biophysical time series where the exact calibration of measurement devices is unknown or varies with time. In real-life applications, such changes in the calibration occur, e.g., for electroencephalography (EEG) recordings where small displacements of electrodes result in changes of the conductivity.

Detecting changes in the dynamics of a time series by looking at ordinal pattern distributions has first been proposed by Bandt and Pompe (2002). More specifically, they consider permutation entropy, which is the Shannon entropy of ordinal pattern distributions, as a measure for detecting changes in the complexity of time series. Permutation entropy has been applied to the analysis of epileptic activity in EEG data (Li et al., 2007; Bruzzo et al., 2008) and to measuring anaesthetic drug effects (Li et al., 2008; Olofsen et al., 2008). In methodological studies, Keller et al. (2007a, 2007b) define further statistics of ordinal pattern distributions besides permutation entropy and investigate properties of the distributions themselves. Bandt and Shiha (2007) derive formulas for ordinal pattern probabilities in stochastic processes; Sinn and Keller (2011) study statistical properties of estimators of these probabilities.

This paper is organized as follows: In Section 2, we formalize the change-point problem and review the definition of ordinal pattern distributions. Section 3 shows how MMD can be used to automatically detect change-points and discusses related work. In Section 4, we evaluate the performance of our methods in simulation studies and demonstrate the application to real-life time series. Section 5 concludes the paper.

# 2 OUTLINE OF THE METHOD

In this paper, we consider the problem of inferring the time point of a change in the distribution of an observed time series. The basic idea is to look at the order structure in different parts of the time series; if we find a time point with a clear difference between the order structure before and afterwards, we conclude that it is a change-point.

Formally, we describe the problem as follows: Let $\mathbb{N}$ denote the set of natural numbers and $\mathbb{Z}$ the set of integers. Consider a real-valued stochastic process $\mathbf{X} = (X_t)_{t\in\mathbb{Z}}$ on some probability space $(\Omega, \mathcal{A}, \mathbb{P})$, where $\Omega$ denotes the sample space, $\mathcal{A}$ the set of measurable events and $\mathbb{P}$ the probability measure. Let $\mathbf{Y} = (Y_t)_{t\in\mathbb{Z}}$ be the process of increments given by $Y_t := X_t - X_{t-1}$ for $t \in \mathbb{Z}$. Throughout this paper, we always assume the following holds:

(A1) The process $\mathbf{Y}$ is non-degenerate, i.e., all finite-dimensional distributions of $\mathbf{Y}$ are absolutely continuous with respect to the Lebesgue measure.

(A2) The processes $(Y_0, Y_{-1}, \ldots)$ and $(Y_1, Y_2, \ldots)$ are strictly stationary and ergodic.

Note that as a consequence of (A1), the values of $\mathbf{X}$ are pairwise distinct $\mathbb{P}$-almost surely, that is,

$$\mathbb{P}(X_{t_1} \neq X_{t_2}) \quad = \quad 1 \tag{1}$$

for all $t_1, t_2 \in \mathbb{Z}$ with $t_1 \neq t_2$.

Now let $m, n \in \mathbb{N}$. Suppose that we observe a time series of length $m+n$ with a change in the underlying distribution after the first $m$ time points, and $m, n$ are unknown (only the length $m+n$ is observable). Then finding the location of the change-point is equivalent to estimating the pair $(m, n)$. We model this situation by assuming the observed time series is a realization of $\mathbf{X}$ at times $t = -m, -m+1, \ldots, n-2, n-1$. If the distributions of $(Y_0, Y_{-1}, \ldots)$ and $(Y_1, Y_2, \ldots)$ are different, then $t = 0$ is a change-point in $\mathbf{X}$ and we obtain $m$ observations before and $n$ observations afterwards.

A generalization of the situation with only one change in the distribution is the multiple change-point problem where the time series potentially has more than one change-point. A common approach to solve this problem is to iteratively apply a method for detecting single change-points, i.e., after the detection of the first change-point, the method is applied to the time segments before and afterwards to search for further change-points. In Section 4.1, we will consider this approach in more detail.

## 2.1 ORDINAL PATTERN DISTRIBUTIONS

We use the concept of ordinal pattern distributions to describe the order structure of a time series. An ordinal pattern represents the order relations among successive values of a time series; if the values are pairwise different, it is natural to identify ordinal patterns with permutations. Counting the number of occurrences of the permutations in a time series (or parts of it) yields the empirical distribution of ordinal patterns.

Here we give a formal definition of ordinal patterns. Let $d \in \mathbb{N}$ be fixed. By $\mathbf{\Pi}$ we denote the set of permutations $\{0, 1, \ldots, d\}$, which we represent by $(d+1)$-tuples containing each of the numbers $0, 1, \ldots, d$ exactly one time. Let the permutations be denoted by $\boldsymbol{\pi}_1, \boldsymbol{\pi}_2, \ldots, \boldsymbol{\pi}_{(d+1)!}$ so that $\mathbf{\Pi} = \{\boldsymbol{\pi}_1, \boldsymbol{\pi}_2, \ldots, \boldsymbol{\pi}_{(d+1)!}\}$. Now consider the partition of $\mathbb{R}^{d+1}$ obtained by identifying vectors whose components are correspondingly ordered: For $\boldsymbol{\pi} = (r_0, r_1, \ldots, r_d) \in \mathbf{\Pi}$, let $B(\boldsymbol{\pi})$ be the subset of $\mathbb{R}^{d+1}$ containing every vector $(x_0, x_1, \ldots, x_d)$ satisfying

($i$) $x_{r_0} \geq x_{r_1} \geq \ldots \geq x_{r_d}$,

($ii$) if $x_{r_i} = x_{r_{i+1}}$, then $r_i > r_{i+1}$.

Obviously, the union of all subsets $B(\boldsymbol{\pi}_1)$, $B(\boldsymbol{\pi}_2)$, $\ldots$, $B(\boldsymbol{\pi}_{(d+1)!})$ is equal to $\mathbb{R}^{d+1}$, and by condition ($ii$) the pair of subsets $B(\boldsymbol{\pi}_i)$, $B(\boldsymbol{\pi}_j)$ is disjoint if $i \neq j$. By saying that the *ordinal pattern* of *order* $d$ at *time* $t \in \mathbb{Z}$ is given by $\boldsymbol{\pi} = (r_0, r_1, \ldots, r_d)$, we designate the event

$$(X_t,\, X_{t-1},\, X_{t-2}, \ldots,\, X_{t-d}) \quad \in \quad B(\boldsymbol{\pi}).$$

According to (1), this event is equivalent to

$$X_{t-r_0} \; > \; X_{t-r_1} \; > \; \ldots \; > \; X_{t-r_d}$$

$\mathbb{P}$-almost surely. Note that a more general definition of ordinal patterns includes an additional parameter $\tau \in \mathbb{N}$ (called the *delay*) which allows us to consider the events $(X_t,\, X_{t-\tau}, \ldots,\, X_{t-d\tau}) \in B(\boldsymbol{\pi})$ (see Keller et al. (2007a)).

Figure 1 shows all the possible outcomes for ordinal patterns of order $d = 3$. For example, if $\omega \in \Omega$ is such that $X_t(\omega) > X_{t-1}(\omega) > X_{t-2}(\omega) > X_{t-3}(\omega)$ (thus, the time series roughly looks like the upper left plot in Figure 1), then the ordinal pattern at time $t$ is given by (0,1,2,3). If $X_t(\omega) > X_{t-1}(\omega) > X_{t-3}(\omega) > X_{t-2}(\omega)$, then the ordinal pattern is given by $(0, 1, 3, 2)$, and so on. The dark gray patterns $(0, 1, 2, 3)$ and $(3, 2, 1, 0)$ in Figure 1 occur if the underlying time series is monotonically increasing and decreasing, respectively. The light gray patterns occur if the past two increments

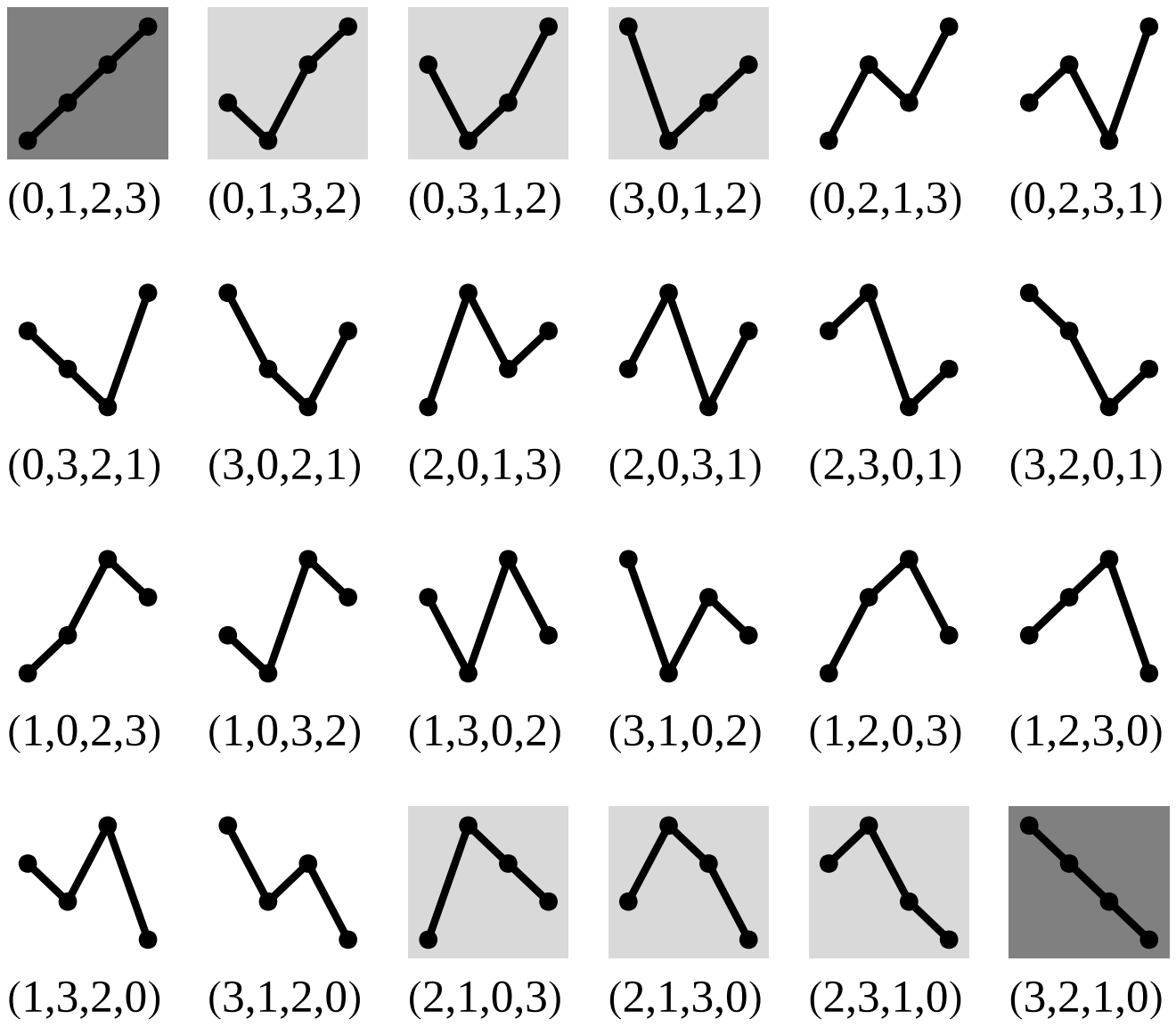

Figure 1: Possible outcomes for ordinal patterns of order $d = 3$.

have the same sign, and the white patterns occur after changes between "upwards" and "downwards".

Next, we consider the distribution of ordinal patterns in the process $\mathbf{X}$. For $t \in \mathbb{Z}$ and $k = 1, 2, \ldots, (d+1)!$ define

$$p_k(t) \quad := \quad \mathbb{P}(\,(X_t,\, X_{t-1},\, \ldots,\, X_{t-d}) \,\in\, B(\boldsymbol{\pi}_k)\,),$$

that is, $p_k(t)$ denotes the probability that the ordinal pattern at time $t$ is given by $\boldsymbol{\pi}_k$. By the *distribution* of ordinal patterns at time $t$ we mean the stochastic vector

$$\mathbf{p}(t) \quad := \quad (p_1(t),\, p_2(t),\, \ldots,\, p_{(d+1)!}(t))\,.$$

Note that the ordinal pattern at time $t$ actually only depends on the increments $Y_t, Y_{t-1}, \ldots, Y_{t-d+1}$ (see Sinn and Keller (2011)). Hence, a sufficient condition for the ordinal pattern distributions $\mathbf{p}(t_1)$ and $\mathbf{p}(t_2)$ to be identical is that the random vectors $(Y_{t_1}, Y_{t_1-1}, \ldots, Y_{t_1-d+1})$ and $(Y_{t_2}, Y_{t_2-1}, \ldots, Y_{t_2-d+1})$ have the same distribution. According to assumption (A2), we obtain that

$$\ldots \,=\, \mathbf{p}(-1) \,=\, \mathbf{p}(0) \quad \text{and} \quad \mathbf{p}(d) \,=\, \mathbf{p}(d+1) \,=\, \ldots,$$

i.e., the ordinal pattern distribution at times $t \leq 0$ and $t \geq d$, respectively, are identical. In general, the distributions $\mathbf{p}(t)$ with $t = 1, 2, \ldots, d-1$ are equal neither to $\mathbf{p}(0)$ nor to $\mathbf{p}(d)$ because they depend on increments both before and after $t = 0$.

In the following, we write $\mathbf{p}_-$ and $\mathbf{p}_+$ to denote the generic distributions $\mathbf{p}(0)$ and $\mathbf{p}(d)$, respectively. Obviously, a necessary condition for $\mathbf{p}_- \neq \mathbf{p}_+$ is that $\mathbf{X}$ has a change-point at $t = 0$. The key idea of our method for change-point detection is to measure the difference between the empirical ordinal pattern distributions in different parts of a time series.

### 2.2 EMPIRICAL ORDINAL PATTERN DISTRIBUTIONS

In the following, let $\mathbf{1}\{\cdot\}$ denote the function which evaluates to 1 if the statement in the brackets is true, and to 0 otherwise. For $w \in \mathbb{N}$ and $t \in \mathbb{Z}$, define

$$\hat{p}_k(w,t) := \frac{1}{w} \sum_{s=wt+1}^{wt+w} \mathbf{1}\big\{\,(X_s,\, X_{s-1},\, \ldots,\, X_{s-d}) \,\in\, B(\boldsymbol{\pi}_k)\,\big\},$$

that is, $\hat{p}_k(w,t)$ is the relative frequency of time points in the time window $\{wt{+}1, wt{+}2, \ldots, wt{+}w\}$ at which the ordinal pattern is $\boldsymbol{\pi}_k$. The parameter $w$ determines the size of the time window; as we will see in Theorem 1 below, the larger $w$, the closer we can expect $\hat{p}_k(w,t)$ be to $p_k(t)$.

Now, taking into account all the estimates $\hat{p}_k(w,t)$ for $k = 1, 2, \ldots, (d+1)!$ gives us the empirical ordinal pattern distribution

$$\hat{\mathbf{p}}(w,t) \quad := \quad \big(\hat{p}_1(w,t),\, \hat{p}_2(w,t),\, \ldots,\, \hat{p}_{(d+1)!}(w,t)\big)\,.$$

As the following theorem shows, $\hat{\mathbf{p}}(w,t)$ converges to $\mathbf{p}_-$ if $t < 0$, and to $\mathbf{p}_+$ if $t \geq 0$ ($\mathbb{P}$-almost surely). This result also holds if the "calibration" of $\mathbf{X}$ changes at a number of points that grows sublinearly in time.

**Theorem 1** *Suppose that conditions* (A1)*and* (A2) *hold. Then*

$$\lim_{w\to\infty} \hat{\mathbf{p}}(w,t) \quad = \quad \begin{cases} \mathbf{p}_- & \text{if } t < 0, \\ \mathbf{p}_+ & \text{if } t \geq 0, \end{cases}$$

$\mathbb{P}$*-almost surely. This result also holds if we observe the "distorted" process* $(h(X_t,t))_{t\in\mathbb{Z}}$ *instead of* $(X_t)_{t\in\mathbb{Z}}$ *with*

$$h(x,t) \quad := \quad \sum_{j\in\mathbb{Z}} \mathbf{1}\{t \in (t_{j-1}, t_j]\}\, h_j(x),$$

*where the mappings* $h_j : \mathbb{R} \to \mathbb{R}$ *are strictly increasing for all* $j \in \mathbb{Z}$*, and* $(t_j)_{j\in\mathbb{Z}}$ *is an increasing sequence in* $\mathbb{Z}$ *which satisfies*

$$\lim_{w\to\infty} \frac{\sharp\{j \in \mathbb{Z} :\ t_j \in [-w,w]\}}{w} \quad = \quad 0.$$

*Proof.* First, we consider the case $t < 0$. Since $(Y_0, Y_{-1}, \ldots)$ is ergodic and

$$\mathbb{E}\big[\,\mathbf{1}\{(X_s,\, X_{s-1},\ldots,\, X_{s-d}) \,\in\, B(\boldsymbol{\pi}_k)\}\,\big] \quad = \quad p_k(0)$$

for all $s \leq w(t+1)$ with $w \in \mathbb{N}$, the Ergodic Theorem shows that $\hat{p}_k(w,t)$ converges to $p_k(0)$ $\mathbb{P}$-almost surely (Cornfeld et al., 1982). If $t > 0$, then $wt + 1 \geq d$ for sufficiently large $w$, and hence

$$\mathbb{E}\big[\,\mathbf{1}\{(X_s,\, X_{s-1},\ldots,\, X_{s-d}) \,\in\, B(\boldsymbol{\pi}_k)\}\,\big] \quad = \quad p_k(d)$$

for all $s \geq wt+1$. Since $(Y_1, Y_2, \ldots)$ is ergodic, the Ergodic Theorem shows that $\hat{p}_k(w,t)$ converges to $p_k(d)$ $\mathbb{P}$-almost surely. In the case $t = 0$, we have

$$\begin{aligned} &\hat{p}_k(w,t) \\ &\quad = \frac{1}{w}\sum_{s=1}^{d-1} \mathbf{1}\{(X_s,\, X_{s-1},\, \ldots,\, X_{s-d}) \,\in\, B(\boldsymbol{\pi}_k)\} \\ &\quad + \frac{1}{w}\sum_{s=d}^{w} \mathbf{1}\,\{(X_s,\, X_{s-1},\, \ldots,\, X_{s-d}) \,\in\, B(\boldsymbol{\pi}_k)\}. \end{aligned}$$

Since the first term on the right hand side tends to 0 and the second one converges to $p_k(d)$ $\mathbb{P}$-almost surely, the result follows.

Now suppose that we observe $(h(X_t,t))_{t\in\mathbb{Z}}$ instead of $\mathbf{X}$. Note that if $t, j \in \mathbb{Z}$ satisfy $t - d \geq t_{j-1}$ and $t \leq t_j$, then $(h(X_t),\, h(X_{t-1}),\, \ldots,\, h(X_{t-d}))$ lies in $B(\boldsymbol{\pi}_k)$ if and only if $(X_t,\, X_{t-1},\, \ldots,\, X_{t-d})$ lies in $B(\boldsymbol{\pi}_k)$. Therefore, for any $t \in \mathbb{Z}$, the number of time points $s \in \{wt+1,\, wt+2,\, \ldots,\, wt+w\}$ at which the indicator functions of these two events are unequal is bounded by $(d+1)$-times the number of $j \in \mathbb{Z}$ for which $t_j \in [wt+1, wt+w]$. Since the latter number divided by $w$ tends to 0 as $w \to \infty$, we obtain the result. □

### 2.3 DETECTING CHANGE-POINTS

Based on the definition of ordinal pattern distributions and the consistency result of Theorem 1, we now present our method for change-point detection. Suppose that $\mathbf{X}$ has a change-point at time $t = 0$ which results in a difference between the ordinal pattern distributions $\mathbf{p}_-$ and $\mathbf{p}_+$. Furthermore, suppose that we observe $\mathbf{X}$ at times $t = -wm+1, -wm+2, \ldots, wn-1, wn$ where $w \in \mathbb{N}$ is "large" and $m, n \in \mathbb{N}$ are unknown (only the sum $m+n$ is observed). In this setting, the problem of estimating the location of the change-point is equivalent to estimating the unknown values $m$ and $n$. Our proposed method proceeds as follows:

1. We partition the observations of $\mathbf{X}$ into $(m+n)$ non-overlapping blocks of size $w$ each.
2. For each block, we compute the empirical ordinal pattern distribution, which gives us the $(m+n)$-dimensional vector

$$\mathbf{z}(w) := \qquad (2)$$
$$(\hat{\mathbf{p}}(w,-m),\, \hat{\mathbf{p}}(w,-m+1),\, \ldots,\, \hat{\mathbf{p}}(w,n-1)).$$

According to Theorem 1, we have

$$\lim_{w\to\infty} \mathbf{z}(w) \quad = \quad (\underbrace{\mathbf{p}_-, \ldots, \mathbf{p}_-}_{m \text{ times}}, \underbrace{\mathbf{p}_+, \ldots, \mathbf{p}_+}_{n \text{ times}}) \qquad (3)$$

$\mathbb{P}$-almost surely. Thus, if the $m+n$ ordinal pattern distributions can be divided into two clearly distinct groups, the sizes of these groups can be taken as estimates for $m$ and $n$. In the following section, we apply Maximum Mean Discrepancy to determine these groups.

Before discussing related work, let us make some remarks:

**Remark 1.** Our method is robust with respect to changes in the "calibration" of the process $\mathbf{X}$. In particular, if $h_-$ and $h_+$ are both strictly increasing functions, then the ordinal pattern distributions $\mathbf{p}_-$ and $\mathbf{p}_+$ in the transformed process $(\ldots,\, h_-(X_{-1}),\, h_-(X_0),\, h_+(X_1),\, h_+(X_2),\, \ldots)$ are identical. As mentioned in the introduction, this robustness is often desirable for the analysis of biophysical time series the exact calibration of which is unknown or varies with time.

**Remark 2.** If $d$ is large enough, then our method does detect changes of the complexity of $\mathbf{X}$ measured by the permutation entropy (Bandt and Pompe, 2002; Keller and Sinn, 2009). Furthermore, it is capable to detect changes in the autocorrelation structure of Gaussian processes (Bandt and Shiha, 2007).

**Remark 3.** There is some trade-off between the choice of the order of the ordinal patterns $d$ and the time window size $w$. In general, a larger order $d$ permits to detect changes of the higher-order autocorrelations of $\mathbf{Y}$ or, equivalently, in the lower frequency domain. On the other hand, the time window size $w$ needs to be much larger than $(d+1)!$ in order to obtain reliable estimates of all the ordinal pattern probabilities, however, the larger $w$, the less accurate the localization of the change-points. For our experiments in Section 4, we have chosen $d = 3$ and $w = 500$, that is, we consider 24 different ordinal patterns and are capable to localize change-points with a precision of $\pm 250$ time points which, for the EEG and ECG recordings, corresponds to approximately $\pm 1$ second. See also Section 3.3 for a further discussion on this issue.

# 3 MAXIMUM MEAN DISCREPANCY

In order to group the components of the vector $\mathbf{z}(w)$ defined in (2), we use a recently proposed criterion called Maximum Mean Discrepancy (MMD). Let $\mathcal{Z}$ be an arbitrary set and $\mathbf{z} = (z_1, z_2, \ldots, z_{m+n}) \in \mathcal{Z}^{m+n}$ with $m, n \in \mathbb{N}$. MMD is a new criterion for testing whether $z_1, z_2, \ldots, z_m$ and $z_{m+1}, z_{m+2}, \ldots, z_{m+n}$ come from different distributions (Gretton et al., 2007a, 2007b). The basic approach is to measure the similarity of points $z, z' \in \mathcal{Z}$ using a kernel $k(z, z')$. In our case, $\mathcal{Z}$ is the set of $(d + 1)!$-dimensional stochastic vectors, and for $k(z, z')$ we choose the Radial Basis Function (RBF) kernel

$$k(z, z') \quad = \quad \exp\left(-\frac{\|z - z'\|^2}{2\sigma^2}\right)$$

where $\|\cdot\|$ denotes the Euclidean norm and $\sigma^2 > 0$ is a smoothing parameter. For $m', n' \in \mathbb{N}$ with $m' + n' = m + n$, the *Maximum Mean Discrepancy* is given by

$$\mathrm{MMD}(\mathbf{z}, m', n') \quad := \\ \left(\frac{\mathrm{K}_1(\mathbf{z}, m', n')}{m'm'} - \frac{2\,\mathrm{K}_2(\mathbf{z}, m', n')}{m'n'} + \frac{\mathrm{K}_3(\mathbf{z}, m', n')}{n'n'}\right)^{\frac{1}{2}}$$

where

$$\mathrm{K}_1(\mathbf{z}, m', n') \quad := \quad \sum_{i=1}^{m'} \sum_{j=1}^{m'} k(z_i, z_j)\,,$$

$$\mathrm{K}_2(\mathbf{z}, m', n') \quad := \quad \sum_{i=1}^{m'} \sum_{j=1}^{n'} k(z_i, z_{m'+j})\,,$$

$$\mathrm{K}_3(\mathbf{z}, m', n') \quad := \quad \sum_{i=1}^{n'} \sum_{j=1}^{n'} k(z_{m'+i}, z_{m'+j})\,.$$

The terms $\frac{1}{m'm'}\mathrm{K}_1(\mathbf{z}, m', n')$ and $\frac{1}{n'n'}\mathrm{K}_3(\mathbf{z}, m', n')$ can be regarded as the average similarity of points within the groups $z_1, z_2, \ldots, z_{m'}$ and $z_{m'+1}, z_{m'+2}, \ldots, z_{m'+n'}$, respectively, while $\frac{1}{m'n'}\mathrm{K}_2(\mathbf{z}, m', n')$ measures the average similarity between the two groups. The maximal MMD value is obtained for that partition of $\mathbf{z}$ which yields maximum similarity within the groups and maximum dissimilarity between them. The next theorem shows how the MMD criterion can be used to locate change-points in $\mathbf{X}$.

**Theorem 2** *Suppose that conditions* (A1) *and* (A2) *hold and let* $\mathbf{z}(w)$ *be as defined in* (2)*. Then for all* $m', n' \in \mathbb{N}$ *with* $m' + n' = m + n$*, we have*

$$\lim_{w\to\infty} \mathrm{MMD}(\mathbf{z}(w), m', n') \\ = \min\left\{\frac{n}{n'}, \frac{m}{m'}\right\} \sqrt{2\left(1 - k(\mathbf{p}_-, \mathbf{p}_+)\right)}$$

$\mathbb{P}$*-almost surely. In particular, if* $\mathbf{p}_- \neq \mathbf{p}_+$*, then*

$$\lim_{w\to\infty} \Big(\arg \max_{\substack{(m',n')\in\mathbb{N}^2,\\ m'+n'=m+n}} \mathrm{MMD}(\mathbf{z}(w), m', n')\Big) = (m, n)$$

$\mathbb{P}$*-almost surely.*

*Proof.* Suppose that $m' \leq m$. By equation (3) and the fact that $k(z, z) = 1$ for all $z \in \mathcal{X}$, the limits of $\mathrm{K}_1(\mathbf{z}(w), m', n')$, $\mathrm{K}_2(\mathbf{z}(w), m', n')$ and $\mathrm{K}_3(\mathbf{z}(w), m', n')$ are given by $m'm'$, $m'(m - m') + m'n\,k(\mathbf{p}_-, \mathbf{p}_+)$ and $n'(n'-2n)+2n(n'-n)\,k(\mathbf{p}_-, \mathbf{p}_+)$, respectively. Putting all terms together, we obtain

$$\lim_{w\to\infty} \mathrm{MMD}(\mathbf{z}(w), m', n') = \frac{n}{n'} \sqrt{2\left(1 - k(\mathbf{p}_-, \mathbf{p}_+)\right)}.$$

In the case $m' > m$, which is equivalent to $n' \leq n$, the result follows by symmetry. Now the second statement is obvious. □

## 3.1 BIAS CORRECTION

Theorem 2 shows that if the ordinal pattern distributions $\mathbf{p}_-$ and $\mathbf{p}_+$ are different, then the argument $(m', n')$ which maximizes $\mathrm{MMD}(\mathbf{z}(w), m', n')$ is a strongly consistent estimator of $(m, n)$. For finite $w$, however, practical experiments suggest that this estimator is biased towards $(1, m+n-1)$ and $(m+n-1, 1)$, particularly when the difference between $\mathbf{p}_-$ and $\mathbf{p}_+$ is small (see Section 4.1). Here we give a heuristic explanation: suppose that the similarity between any two components of $\mathbf{z} = (z_1, z_2, \ldots, z_{m+n})$, as measured by the RBF kernel, is approximately equal to some constant $\delta > 0$. For the random vector $\mathbf{z}(w)$, this is likely to be the case if the difference between $\mathbf{p}_-$ and $\mathbf{p}_+$ is so small that all estimates are scattered in the same region. A simple calculation shows that

$$\mathrm{MMD}(\mathbf{z}, m', n') \quad \approx \quad \left(1 - \delta\right) \frac{m + n}{m'n'} \qquad (4)$$

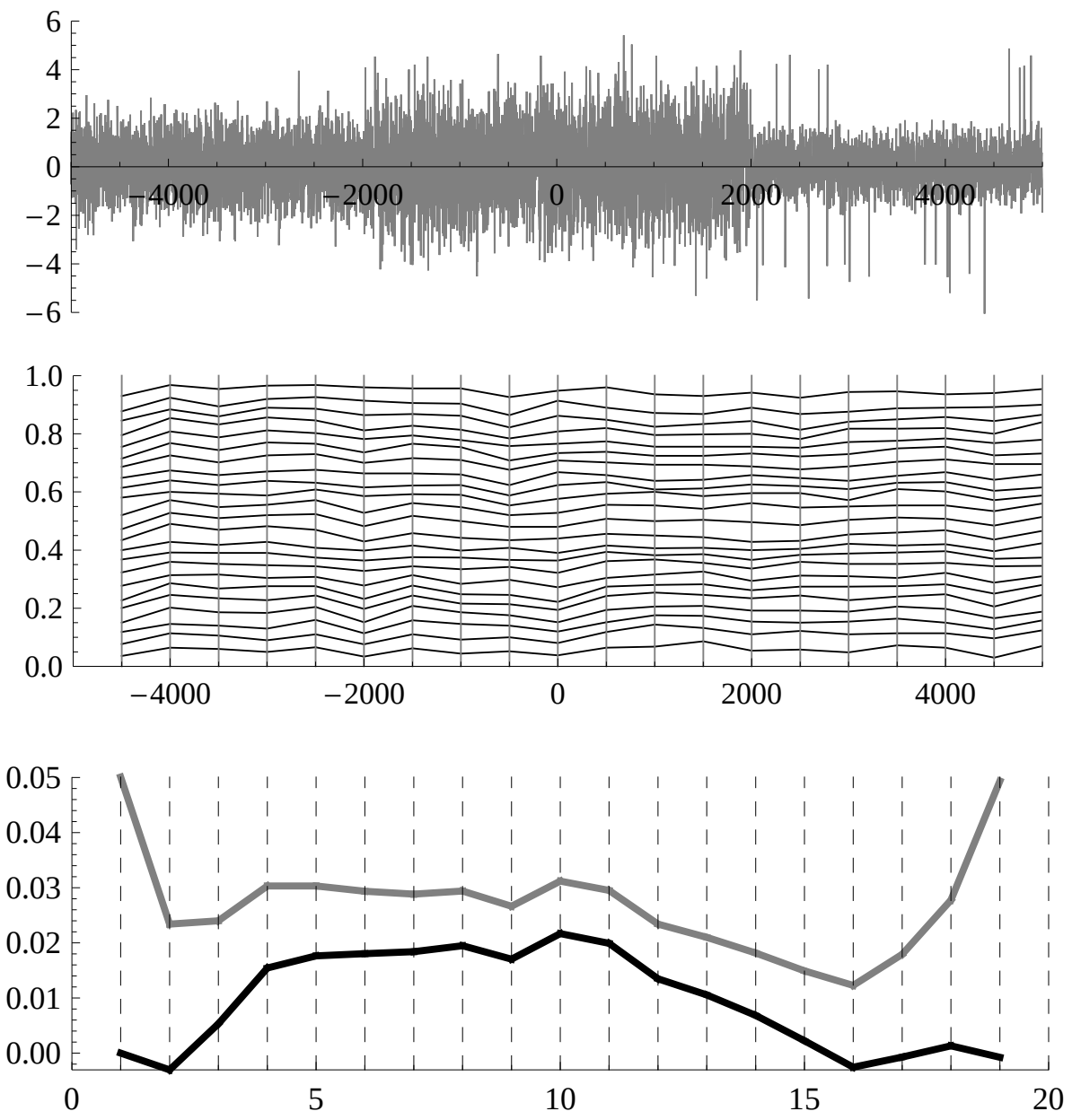

Figure 2: Top: Realization of an AR(1) process with a change-point at $t = 0$. Middle: Ordinal pattern distributions in time windows of size $w = 500$. Bottom: Resulting values for MMD (gray) and CMMD (black).

for $m', n' \in \mathbb{N}$ with $m' + n' = m + n$, and hence the argument maximizing $\mathrm{MMD}(\mathbf{z}, m', n')$ is likely to be $(m', n') = (1, m + n - 1)$ or $(m', n') = (m + n - 1, 1)$. Next we propose a correction for the bias: define the *Corrected Maximum Mean Discrepancy*

$$\mathrm{CMMD}(\mathbf{z}, m', n') := \mathrm{MMD}(\mathbf{z}, m', n') - \frac{m+n-1}{m'n'} \max_{\substack{(\tilde{m},\tilde{n})\in\mathbb{N}^2,\\ \tilde{m}+\tilde{n}=m+n}} \mathrm{MMD}(\mathbf{z}, \tilde{m}, \tilde{n}).$$

The correction term on the right hand side can be regarded as an estimate of the bias of the MMD statistic (compare to (4)). Experimental results show the superiority of using CMMD instead of MMD (see Section 4.1). Note that the estimator of $(m, n)$ based on CMMD is no longer consistent; in order to preserve consistency, the correction term would have to be scaled by a factor that tends to 0 as $w \to \infty$.

### 3.2 EFFICIENT COMPUTATION

Let us discuss the computational complexity of our method. The computation of the ordinal pattern distributions is linear in the length of the input sequence and can be realized by very efficient algorithms (Keller et al., 2007b). Computing the estimate of $(m, n)$ by directly evaluating $\mathrm{MMD}(\mathbf{z}(w), m', n')$ for all $m', n' \in \mathbb{N}$ with $m' + n' = m + n$ requires $O((m + n)^3)$ arithmetic operations (Gretton et al., 2007a). For a more efficient

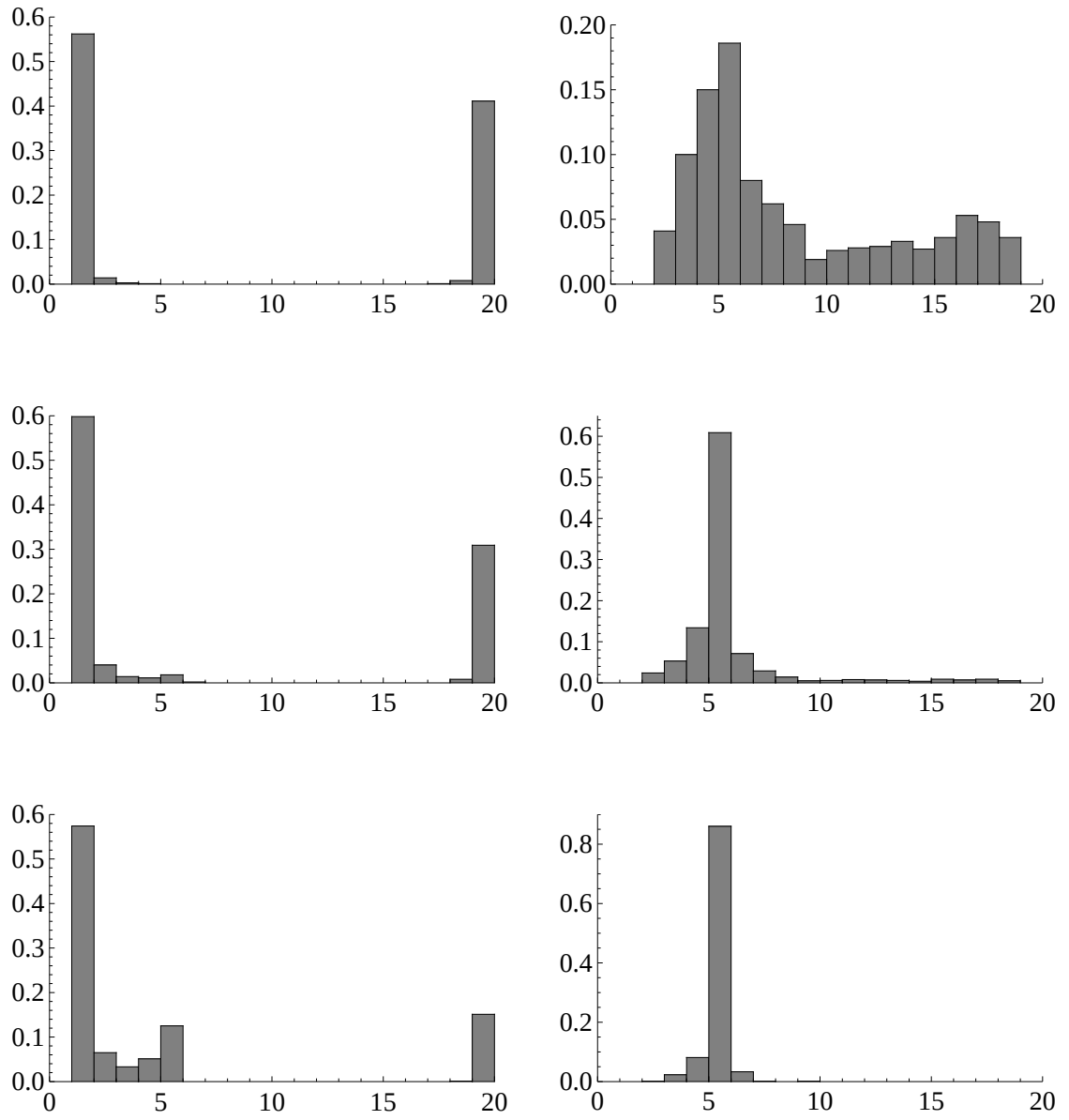

Figure 3: Distribution of the change-point estimators based on MMD (left) and CMMD (right).

computation, note that

$$\begin{aligned}
\mathrm{K}_1(\mathbf{z}, m'+1, n'-1) &= \mathrm{K}_1(\mathbf{z}, m', n') \\
&\quad + 2\sum_{i=1}^{m'} k(z_i, z_{m'+1}) + k(z_{m'+1}, z_{m'+1}), \\
\mathrm{K}_2(\mathbf{z}, m'+1, n'-1) &= \mathrm{K}_2(\mathbf{z}, m', n') \\
&\quad - \sum_{i=1}^{m'} k(z_i, z_{m'+1}) + \sum_{i=1}^{n'-1} k(z_{m'+1}, z_{m'+1+i}), \\
\mathrm{K}_3(\mathbf{z}, m'+1, n'-1) &= \mathrm{K}_3(\mathbf{z}, m', n') \\
&\quad - 2\sum_{i=1}^{n'} k(z_{m'+1}, z_{m'+i}) + k(z_{m'+1}, z_{m'+1}).
\end{aligned}$$

Using these formulas, $\mathrm{MMD}(\mathbf{z}(w), m', n')$ can be evaluated iteratively for all $(m', n') = (1, m + n - 1), \ldots, (m + n - 1, 1)$. Thus, computing the argument which maximizes $\mathrm{MMD}(\mathbf{z}(w), m', n')$ requires $O((m + n)^2)$ arithmetic operations.

### 3.3 RELATED WORK

There is an extensive literature on change-point detection, including both parameteric and non-parametric approaches (Chen and Gupta, 2000; Brodsky and Darkhovsky, 1993, 2000). Recently, a kernel-based method been proposed by Harchaoui et al. (2009). The main difference to our approach is that all these methods aim to find the *exact* location of change-points and give some guarantees on the significance level.

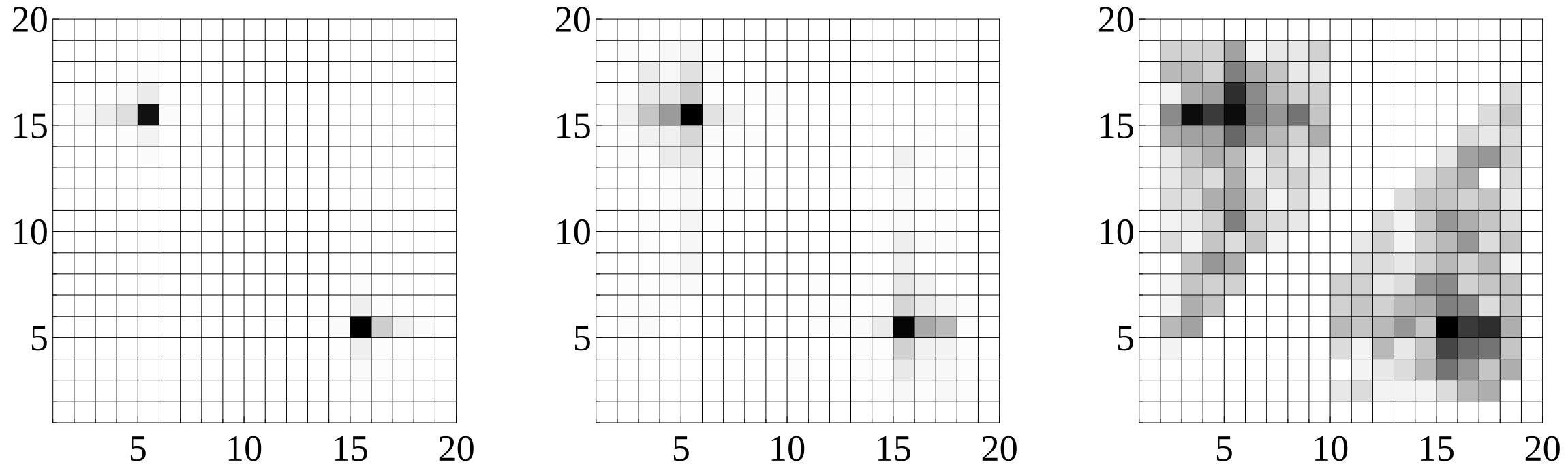

Figure 4: Detection of multiple change-points. The black coloring corresponds to the maximum frequencies 34.0% (left), 15.1% (middle) and 2.2% (right), the white coloring corresponds to 0%.

This precision comes at the price of a high computational burden; for example, Harchaoui's method has a running time which is cubic in the input size. Our method is designed to find the *approximate* location of change-points, where the precision is controlled by the time window size $w$. Because of its computational efficiency, our method is particularly attractive for the fast exploration of high-resolution time series such as EEG recordings, which often consist of hundreds of thousands of data points. Note that our approach can be combined with existing methods in order to first obtain the approximate locations of a set of candidate change-points before determining their exact locations and testing for significance.

# 4 EXPERIMENTS

## 4.1 SIMULATED DATA

**Change of the autoregressive coefficient.** Let us first illustrate our method for simulated data. The top plot of Figure 2 shows a realization of a first order autoregressive (AR(1)) process with standard normal innovations where, at time $t = 0$, the autoregressive coefficient changes from 0.1 to 0.3. There are two changes of the process "calibration": at time $t = -2000$, the variance increases from 1 to 2, and at time $t = 2000$, the variance decreases to $\frac{1}{2}$ and values greater than 2 and smaller than $-2$, respectively, are scaled by the factor 2. According to Remark 1 at the end of Section 2, we are not interested in detecting these changes of the calibration, but only in the change-point at $t = 0$.

The middle plot of Figure 2 shows the distribution of ordinal patterns in time windows of size $w = 500$. The order is $d = 3$, so there are 24 different patterns (compare to Figure 1). Their relative frequencies are represented by the space between horizontal lines: the space between the bottom line and the first line from below represents the relative frequency of the ordinal pattern (0,1,2,3) (the first pattern in Figure 1), the space between the first and the second line from below represents the relative frequency of (0,1,3,2) (the second pattern in Figure 1), and so on. In the framework of Section 2, we have $m = n = 10$, i.e., there are 10 time windows before and after the change-point. As can be seen, the distributions before and after the change-point are slightly different. The bottom plot of Figure 2 shows the resulting MMD and CMMD values for $(m', n') = (1, 19), (2, 18), \ldots, (19, 1)$. In all our experiments, we used $\sigma^2 = 1$ in the RBF kernel, however, we found that the results were not very sensitive with respect to the choice of $\sigma^2$. The maximal MMD value is obtained for $(m', n') = (19, 1)$, while maximizing the argument of CMMD yields the true value $(10, 10)$.

**Comparing MMD and CMMD.** In Figure 3, we compare the distribution of the change-point estimators based on MMD and CMMD. Again, the underlying model is an AR(1) process of length $10,000$ and the time window size is $w = 500$. In this experiment, we chose $(m, n) = (5, 15)$, and the autoregressive coefficient changes from 0.1 to 0.2 (top), from 0.1 to 0.3 (middle) and from 0.1 to 0.4 (bottom). The distributions of the estimators are obtained by $1,000$ Monte Carlo replications. As can be seen, the estimator based on CMMD performs better in all three cases; the one based on MMD is biased towards (1,19) and (19,1), particularly if the change of the autoregressive coefficient is small.

**Multiple change-point detection.** Figure 4 shows the results obtained for the detection of multiple change-points. The underlying model is an AR(1) process of length $10,000$ with a change of the autoregressive coefficient at $2,500$ and $7,500$ time points, namely, from 0.1 to 0.4 and 0.4 to 0.1 (left), from 0.1 to 0.3 and 0.3 to 0.1 (middle), and from 0.1 to 0.2 and 0.2 to 0.1 (right). To localize these change-points, we applied our method twice: first to the entire sequence, then to the longer one of the resulting two time segments. Figure 4 displays the distribution of the estimates obtained by $1,000$ Monte Carlo replications. The coloring of the cell $(x, y)$ represents the frequency of replications with the first change-point detected at $500x$ time points and the second one at $500y$ time points. As can be seen, the localization in the first two experiments is fairly precise. In the third experiment, the estimates are more scattered, however, this is true even in the single change-point setting (compare to Figure 3).

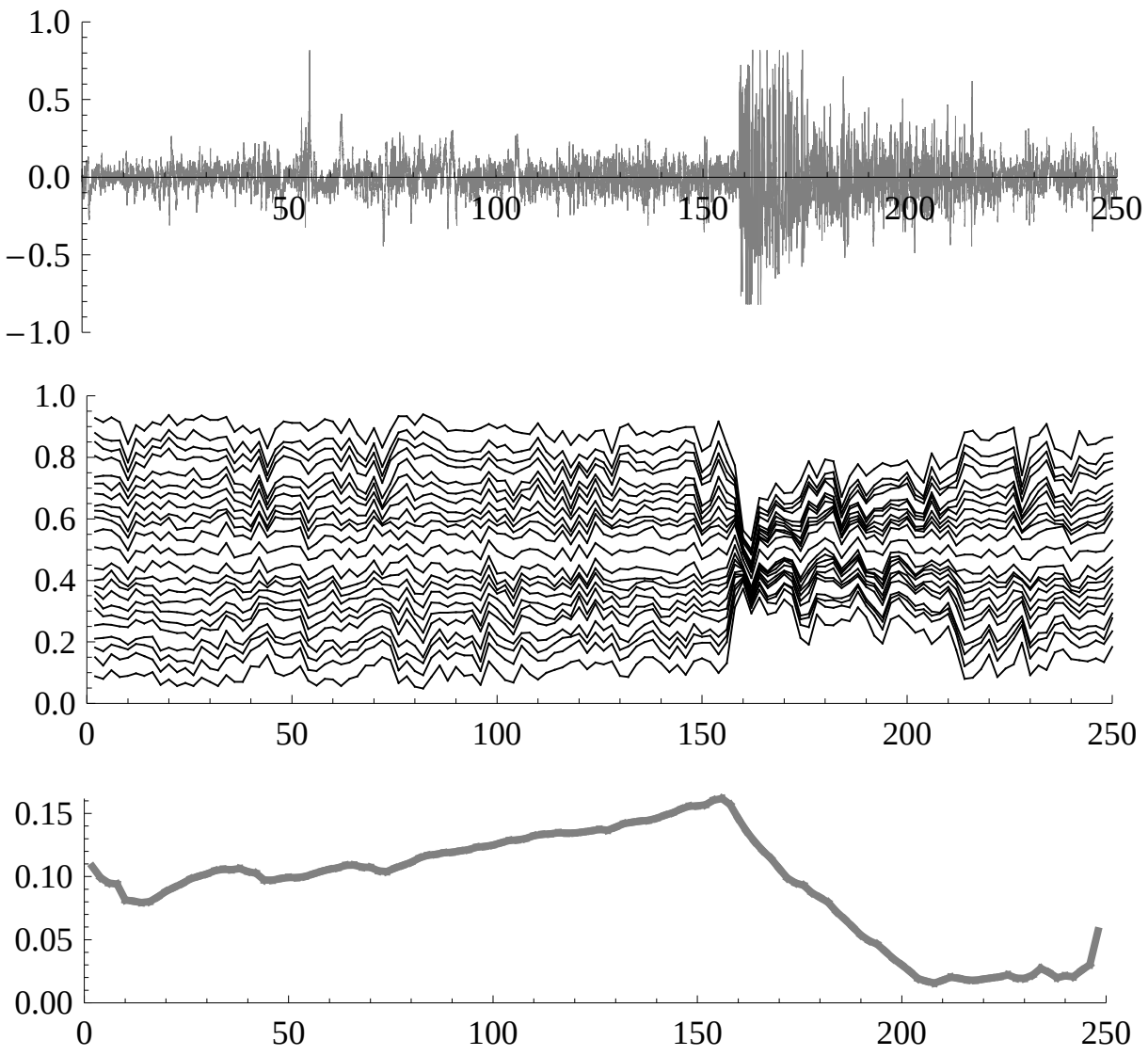

Figure 5: Top: 250 seconds long part of an EEG recording. Middle: Ordinal pattern distributions. Bottom: Resulting MMD values.

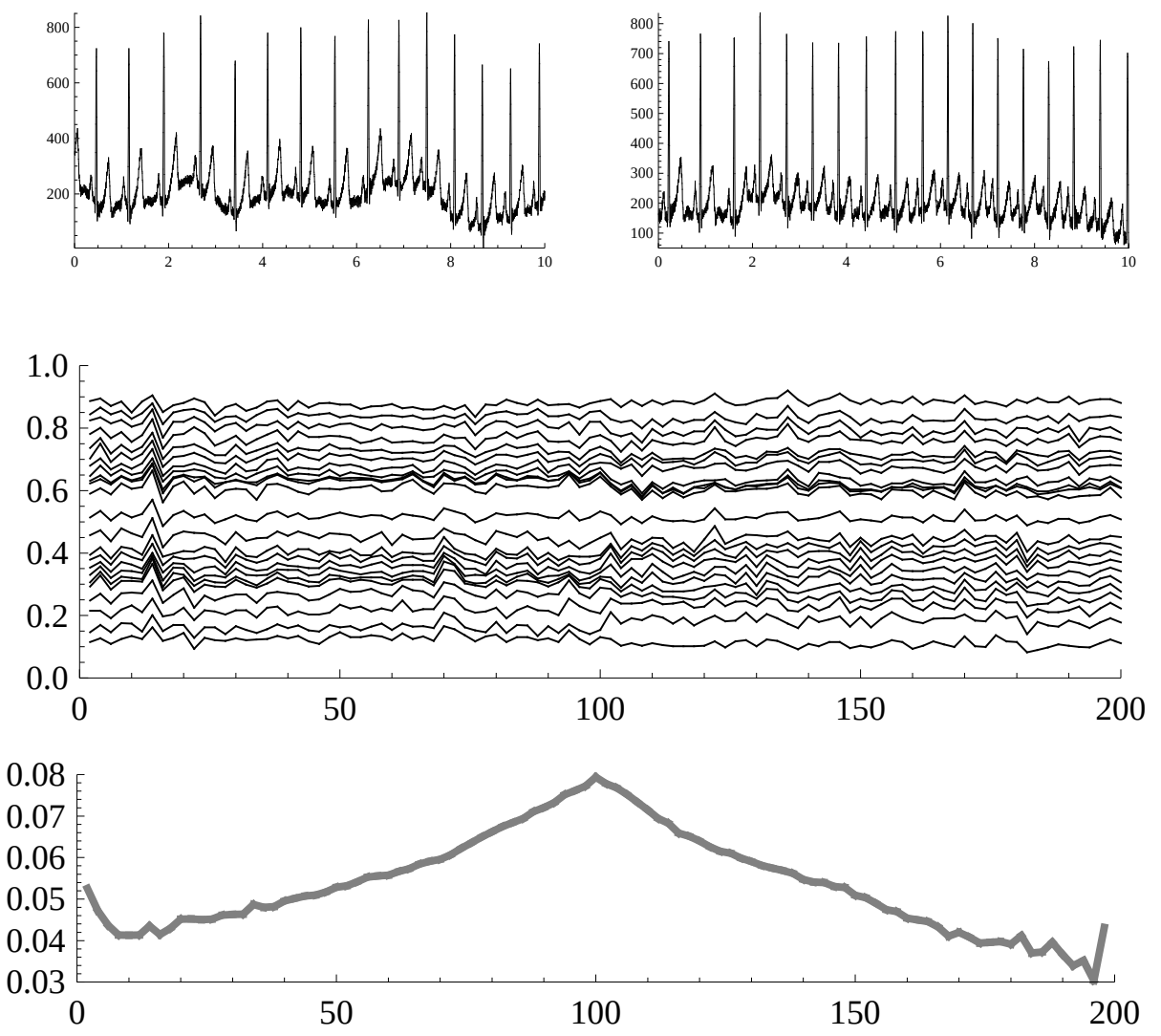

Figure 6: Top: 10 seconds of an ECG before and after an increase of the heart rate. Middle: Ordinal pattern distributions. Bottom: Resulting MMD values.

### 4.2 REAL-LIFE TIME SERIES

**Epileptic activity in EEG time series.** The top plot of Figure 5 shows a 250 seconds long part of an EEG recording which was digitized with a sampling rate of 256 Hertz, so the time series consists of $250 \cdot 256 = 64,000$ data points. The increase in amplitude after 160 seconds is related to the onset of an epileptic seizure. The middle plot shows the distribution of ordinal patterns in non-overlapping time windows of 2 seconds, that is, $w = 512$. Same as in Figure 2, the order of the patterns is $d = 3$. The bottom plot displays the resulting MMD values. The maximum is obtained at the onset of the seizure, thus indicating the presence of a change-point after 160 seconds.

**Heart rate changes in ECG data.** Figure 6 shows the application of our method to the detection of heart rate changes in electrocardiography (ECG) recordings. The middle plot displays the distribution of ordinal patterns in 100 non-overlapping time windows of 2 seconds, again corresponding to the time window size $w = 512$. After 100 seconds, there is a slight change in the heart rate from approximately 90 to 105 beats per minute (compare to the plots in the top row). As the bottom plot shows, this change-point is easily detected even by the non-corrected MMD values.

## 5 CONCLUSIONS

We have presented a new method for detecting change-point in time series based on measuring the maximum mean discrepancy of ordinal pattern distributions. The main advantages of our approach are its computational efficiency and robustness towards (non-linear) signal distortions. These properties make it particularly attractive for the fast exploration of biophysical time series the exact calibration of which is unknown or varies with time. We believe that our approach can be efficiently combined with existing change-point detection methods to first approximately localize of a set of candidates before determining the exact location and testing for significance. We have established asymptotical properties of our method and evaluated its performance both for simulated and real-life data, including the application to multiple change-point problems.